\documentclass[showpacs,twocolumn,prb]{revtex4-1}
\usepackage{graphicx}
\usepackage{subfloat}

\newcommand{\be}{\begin{equation}}
\newcommand{\ee}{\end{equation}}
\newcommand{\bea}{\begin{eqnarray}}
\newcommand{\eea}{\end{eqnarray}}
\newcommand{\bwt}{\begin{widetext}}
\newcommand{\ewt}{\end{widetext}}

\begin{document}

\title{The extended vs standard Holstein model; results in two and three dimensions}

\author{Carl J. Chandler and F. Marsiglio}
\affiliation{Department of Physics, University of Alberta, Edmonton, Alberta, Canada, T6G~2G7}

\begin{abstract}
We present numerically exact solutions to the problem of a single electron interacting through
a long range interaction with optical phonons
in two and three dimensions. Comparisons are made with results for the standard Holstein model,
and with perturbative approaches from both the weak coupling and strong coupling sides. We find,
in agreement with earlier work, that the polaron effective mass increases (decreases) in the weak
(strong) coupling regime, respectively. However, in two dimensions, the decrease in effective mass
still results in too large an effective mass to be relevant in realistic models of normal metals. In three
dimensions the decrease can be more relevant, but exists only over a very limited range of coupling strengths.
\end{abstract}

\date{\today }
\maketitle

\section{introduction}

The standard theoretical framework for superconductivity is known as BCS-Eliashberg
theory,\cite{bardeen57,eliashberg60,scalapino69,allen82,carbotte90,marsiglio08} and, as catalogued in
the cited reviews, accurately describes many experimentally known properties of the so-called 
conventional superconductors, like Pb. BCS theory itself is almost universal, and confirmation of this theory,
achieved with unprecedented accuracy for weak coupling superconductors like Aluminium, serves to vindicate
the``pairing formalism'', utilized to construct BCS theory, but does little to confirm the mechanism.\cite{remark1}

The mechanism for pairing in conventional superconductors is believed to be virtual phonon exchange, in
complete analogy to the virtual photon exchange which is responsible for the direct Coulomb interaction between
two charged particles. The primary evidence for this belief comes from a comparison of tunnelling data with
the {\em deviations} from BCS theory captured in Eliashberg theory, and again, a considerable body of evidence
that confirms the virtual phonon exchange mechanism for pairing is described in 
Refs. [\onlinecite{bardeen57,eliashberg60,scalapino69,allen82,carbotte90,marsiglio08}].

At a more microscopic level, for the past several decades the Holstein model\cite{holstein59,alexandrov07} has 
served as the chief paradigm to describe electron-phonon interactions in solids. In part this paradigm choice has been driven
by the physics, and the realization that in the single polaron problem the interaction can be very local
and the (optical) phonons are well-described by Einstein oscillators. In addition, however, computational
techniques for understanding the properties of a polaron have evolved in a manner conducive to lattice
models with local interactions; this has led to an abundance of studies of the properties of this particular model. Many of these
properties are at odds with the Eliashberg description; early work\cite{scalettar89,marsiglio90} using Quantum Monte
Carlo (QMC) methods suggested a dominance
of charge ordering phenomena in lieu of superconductivity, while more recent work relying on hybrid Migdal-Eliashberg
and Dynamical Mean Field Theory (DMFT) \cite{bauer11} provided some reconciliation, though the competitive charge-ordered 
phase was not included. In any event a programme that begins with a complete understanding of the basic building block
for the many-body state, i.e. of the properties of a single electron coupled to phonons, has been at odds with the Migdal description,
insofar as strongly polaronic properties ensue, even with modest electron-phonon 
coupling,\cite{alexandrov94,marsiglio95,bonca99,fehske07}
particularly when the phonon frequency is small compared to the electronic bandwidth.\cite{li10,alvermann10,remark2}

Alongside these developments the Fr\"ohlich model\cite{frohlich37,frohlich63,devreese07} for 
electron-phonon interactions describes a screened but long-range interaction between an electron and the (positively charged)
ions in a crystal. In fact, it is for this model that much of the early analytical work on the polaron was 
done.\cite{feynman55,kuper63,peeters85}
This model has only one energy scale, the phonon frequency, which makes a comparison with the Holstein model, for example,
difficult. In the Holstein model and other lattice models like it, there are two energy scales, one corresponding to the phonon
frequency and the other corresponding to the (bare) electron bandwidth. Polaronic effects depend significantly on the ratio of
these two energy scales, $\omega_{\rm ph}/t$, sometimes known as the adiabatic ratio. Here, $t$ is the bare electron
hopping amplitude. Until recently,\cite{li10,alvermann10} much of the work done on microscopic models uses an 
adiabatic ratio close to unity; the more physical regime, and the one required by the Migdal approximation when 
many electrons are considered, is $\omega_{\rm ph}/t << 1$.

In an effort to draw comparisons between the short-range Holstein model and the longer-range Fr\"ohlich model, Alexandrov
and Kornilovitch\cite{alexandrov99} defined a Fr\"ohlich polaron problem on a discrete lattice. They examined the behaviour of the
effective mass as a function of coupling strength, primarily on one and two dimensional lattices. They concluded that with
extended range interactions the effective mass can be much smaller than for the Holstein polaron. Thus, a microscopic
model with long-range electron-phonon interactions is a possible means of reconciling exact single electron ``building-block''
calculations with the Migdal approximation that underlies the Eliashberg theory of electron-phonon-mediated superconductivity. 

However, as mentioned earlier, the single electron longer-range interaction studies were carried out with an adiabatic ratio 
of order unity. Here we wish to re-examine this problem with more physical values of the adiabatic ratio, and extend their 
calculations\cite{alexandrov99} to three dimensions. We find that while their conclusion that the effective mass can be much smaller 
than for the Holstein model is correct, this statement applies for a very restrictive range of the coupling strength. We also 
note that the behaviour in two dimensions is not representative of what occurs in either one or three dimensions. In fact, even
in the perturbative regime, low order perturbation theory is not very accurate.
Their initial conclusions are actually more representative for the three dimensional case.

The rest of the paper is as follows. First, following Ref. [\onlinecite{alexandrov99}], we define the model, and we outline the 
method of solution. We have exact results for all our calculations, based on refinements of the method introduced by 
Bon\v ca et al.\cite{bonca99}
This controlled method of solution becomes somewhat more difficult for three dimensions, but we present converged results
for phonon frequencies as low as $\omega_E/t = 0.3$. Considering that this is achieved in three dimensions, where the electronic
bandwidth is $W \equiv 12t$, this phonon energy scale represents $2.5$\% of the electronic bandwidth.

Results are first presented in two dimensions. We first re-assess some older results,\cite{li10} and note that, even for the
standard Holstein model, perturbative calculations, either in weak or strong coupling, are actually not very accurate. In
weak coupling for example, multi-phonon excitations lead to a significantly enhanced effective mass. This phenomenon is
amplified when longer range interactions are included, so in fact we find the conclusions of Ref. [\onlinecite{alexandrov99}]
somewhat misleading. The effective mass does decrease due to longer range interactions in the strong coupling regime,
but in two dimensions, the resulting effective mass is still much too high to be relevant for normal (i.e. non-polaronic) metals.

The following section treats the three dimensional case, where we find in fact that a lower effective mass, {\it to realistic
values}, is indeed achieved by including longer range interactions. However, even here the range of coupling strengths
over which this is achieved is very narrow; in terms of the dimensionless coupling constant $\lambda$ (to be defined below),
this range is very close to unity, and not in the range associated with so-called Eliashberg strong coupling superconductors.
We conclude with a Summary.  

\section{Model and Method} 
 
The lattice Fr\"ohlich (``extended Holstein'') model is defined as\cite{alexandrov99}
\begin{eqnarray}
H = &-& t \sum_{i,\delta} \bigl[c_i^{\dagger} c_{i+\delta} + c_{i+\delta}^{\dagger} c_i \bigr] \nonumber \\
&-& g \omega_E \sum_{\langle \mathbf{i},\mathbf{j} \rangle }  f (\mathbf{j}) n_{\mathbf{i}}  ( a_{\mathbf{i}+\mathbf{j}} + a_{\mathbf{i}+\mathbf{j}}^{\dagger} ) \nonumber \\
&+& \omega_E \sum_i a^{\dagger}_i a_i,
\label{ham1}
\end{eqnarray}
\noindent where the range of the interaction is given by 
\begin{equation}
 f(\mathbf{j}) = \frac{1}{ (\arrowvert \mathbf{j} \arrowvert^2 + 1)^{3/2} }.
 \label{range}
\end{equation}
\noindent In Eq. (\ref{ham1}) $t$ is the electron hopping parameter, $\omega_E$ is the characteristic phonon frequency,
taken to be a constant here, and $g \omega_E f({\mathbf{j}})$ is the coupling strength between an electron at a
particular site and a phonon at a site a distance $a_0|{\mathbf{j}}|$ away, where $a_0$ is the lattice spacing (taking to be unity
hereafter) and $\mathbf{j}$ is the vector connecting the electron and phonon. The case 
$f(\mathbf{j}) = \delta_{\mathbf{j},\mathbf{0}}$ reduces to the usual Holstein model. The other symbols are defined as follows:
$c_i^{\dagger}$ ($c_i$) creates (annihilates) an electron at site i, and $a_i^{\dagger}$ ($a_i$) creates (annihilates) a 
phonon at site i. The electron number operator is given by $n_i \equiv c_i^{\dagger} c_i$, and the spin index has been
suppressed since we are dealing with only one electron. Note that the sum over $\delta$ in the electron hopping part
of the Hamiltonian is over nearest-neighbour sites on the positive side only, to avoid double counting, whereas the sum
over $\mathbf{j}$ in the interaction term in principle includes the on-site term ($\mathbf{j} = 0$ along with neighbouring
sites in {\it all} directions. In practice, following Ref. [\onlinecite{alexandrov99}] we will terminate the sum at nearest neighbour
interactions.  

We will use Eq. (\ref{range}) for all dimensions. In reality this form is motivated by the three dimensional case, where
the long range interaction follows a $1/r^3$ decay. At short distances a potential divergence is cutoff by the constant `1'
in the denominator of Eq. (\ref{range}); this corresponds to a characteristic decay length of the lattice spacing, and in
principle this can be varied as well. Here, for simplicity, we keep the constant fixed at unity. This Hamiltonian contains a dimensionless coupling constant, namely $g$, which becomes most important in the strong coupling regime. 
In practice we also define another dimensionless coupling constant, $\lambda$, which becomes important in the 
weak coupling regime. 

Specifically for the coupling of the electron to the phonons at the same
site we use, following Li et al.,\cite{li10} $\lambda_c \equiv \omega_E g^2/(W/2)$
in one and three dimensions, where $W$ is the bare electron bandwidth, $W = 4t$ in 1D and $W = 12t$ in 3D for the
tight binding model on a linear or cubic lattice with nearest neighbour hopping, respectively.  In two dimensions we use
a definition where the electron density of states at the bottom of the bare band is used [$\equiv 1(4\pi t)$] instead of the average value of the density of states across the entire band [$\equiv 1/(8t)$], so $\lambda_c \equiv \omega_E g^2/(2 \pi t)$. For the
Holstein model the entire coupling would be that of the on-site coupling.

To define the coupling strength for the extended model, $\lambda_{\rm tot}$, we follow the definitions of Alexandrov
and Kornilovitch,\cite{alexandrov99} with the total coupling being the sum of the couplings to the different sites. For example,
in two dimensions we obtain the on-site contribution along with four equally weighted nearest neighbour contributions, reduced
by $[1/2^{3/2}]^2 = 1/8$ compared to the on-site value:
\begin{equation}
 \lambda_{\rm tot} = \sum_{\mathbf{j} } \lambda_{c} f^2(\mathbf{j}) = \lambda_{c}\left( 1 + 4 \frac{1}{8} \right).
\end{equation}

The single polaron problem is solved here with the variational exact diagonalization method described 
in Bon\v ca et al\cite{bonca99} with the same refinements for 
low frequency calculations as developed by Li et al.\cite{li10} The modifications required for the extended model fit 
nicely into this computational framework, with the cost of a denser Hamiltonian matrix, and a Hilbert space that 
grows faster with each application of Hamiltonian, compared to similar calculations for the standard (i.e. on-site) Holstein
model. This rapid growth makes it difficult to converge results using an extended version of the adaptive method of Li et al. 

We therefore further refined the method by producing a list of the most important basis states for each point in 
parameter space.  Starting with a list of basis states from a nearby parameter point previously diagonalized, 
or a truncated coherent state from the strong coupling Lang-Firsov solution\cite{lang63,li10} we diagonalized the 
Hamiltonian in this basis. These basis states in turn were ranked according to the magnitude of their contribution 
to the ground state, and the top $N_1$ contributions were kept. We then acted on these $N_1$ states with the 
Hamiltonian to produce more basis states and diagonalized the Hamiltonian in this new space. The resulting eigenvector
was then sorted, the top $N_1$ contributions were kept, and the process repeated. Once this procedure converged, 
we sorted one last time, and kept the top $N_2$ states ( $N_2 > N_1$) and did the final diagonalization. While this was a
time-consuming calculation, it allowed for much better results with a finite amount of computer memory since 
it selected out the basis states that were the most important for describing the ground state. 

One limit of this method should reduce to weak coupling 2nd order perturbation theory. 
Using straightforward Rayleigh-Schr\"{o}dinger perturbation theory in 2D with on-site and
nearest-neighbour interactions only results in a second-order correction to the ground state energy,
\begin{equation}
E^{(2)}(k_x,k_y) = -{2 \pi \lambda_{\rm tot} t \omega_E \over 1 + 4f^2(1)}{1 \over N}\sum_{k^\prime}
{\biggl(1 - {f(1) \over t}\epsilon_{k^\prime - k}\biggr)^2 \over  \epsilon_{k^\prime} + \omega_E - \epsilon_k},
\label{rs_pert}
\end{equation}
\noindent where $\epsilon_k \equiv -2t[{\rm cos}(k_xa_0) + {\rm cos}(k_ya_0)]$ is the bare energy for the nearest-neighbour
tight-binding model. This expression can be evaluated in terms of complete elliptic integrals --- for example [$f_1 \equiv f(1)$],
\begin{eqnarray}
&&E^{(2)}(0,0) = \nonumber \\
&&-\lambda_{\rm tot} \omega_E \biggl\{ {(1 + 4z f_1)^2 \over 1 + 4f^2_1}{1 \over z} K\bigl[{1 \over z^2}\bigr]
-4\pi f_1 {1 + 2zf_1 \over 1 + 4f^2_1} \biggr\},
\label{rs_per_anal}
\end{eqnarray}
\noindent where $z \equiv 1 + \omega_E/(4t)$, but for most of our perturbation results we have simply evaluated 
the effective mass numerically. Results are shown
for the effective mass in the $k_x$ direction although of course there is a $k_x - k_y$ symmetry.
 
\section{Results in 2D} 

\begin{figure}
\begin{center}
\includegraphics[height=3.7in,width=3.7in]{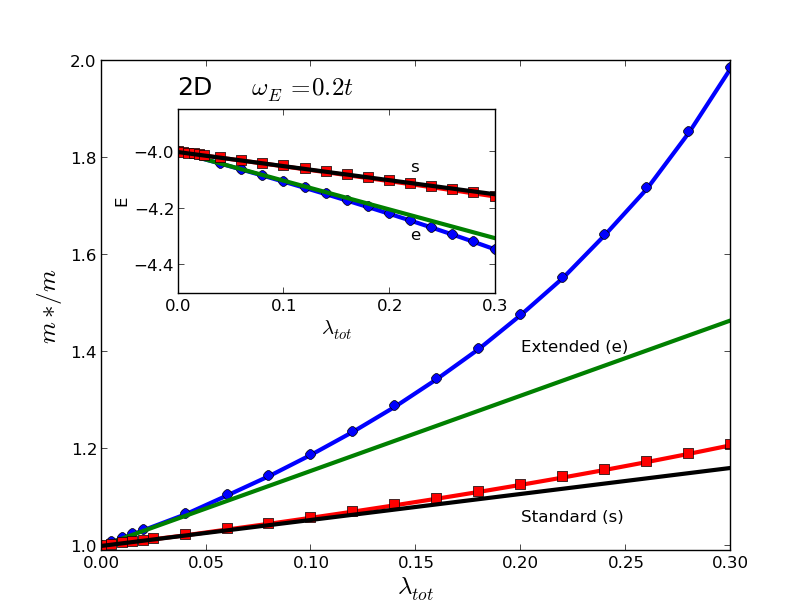}
\caption{ Effective mass, $m^\ast/m$ vs. coupling strength $\lambda_{\rm tot}$, using perturbation theory for both 
the extended and standard Holstein models; these are compared
with results from exact diagonalization, We have used $\omega_E = 0.2t$. Perturbation theory is less accurate 
for the extended model compared with the standard model. {\it Inset:} Ground state energy vs. $\lambda_{\rm tot}$. Note
that the results for the energy are fairly accurate, in comparison to those for the effective mass.}
\label{fig:meff_is_bad_preturbative_2D}
\end{center}
\end{figure}

While others have studied the standard Holstein model in detail,\cite{bonca99,li10,alvermann10} there are a 
few results here that should be emphasized, and which are important for a more complete understanding of the extended Holstein 
model. The perturbative regime for the 2D Holstein model is actually very small. While the ground state energy from perturbation theory matches the exact ground state energy well, the wavefunction and effective mass do not, as 
seen in Fig.~\ref{fig:meff_is_bad_preturbative_2D}. Perturbation 
theory does not work very well because the wavefunction, even at weak coupling, needs to include states with 
multiple phonons, {\it especially} when the phonon frequency is small compared to the electron hopping parameter, $t$. 
Even for the standard Holstein model, restricting our exact diagonalizations to a subspace with a limited number of phonons 
gives quantitatively inaccurate results for the effective mass, even if the energy was not strongly affected, as illustrated in 
Fig.~\ref{fig:meff_N_phonon_2D_ordinary}. For the extended model, the discrepancies are even more pronounced, as 
shown in Fig.~\ref{fig:meff_N_phonon_2D}.

\begin{subfigures}
\begin{figure}
\begin{center}
\includegraphics[height=3.5in,width=3.5in]{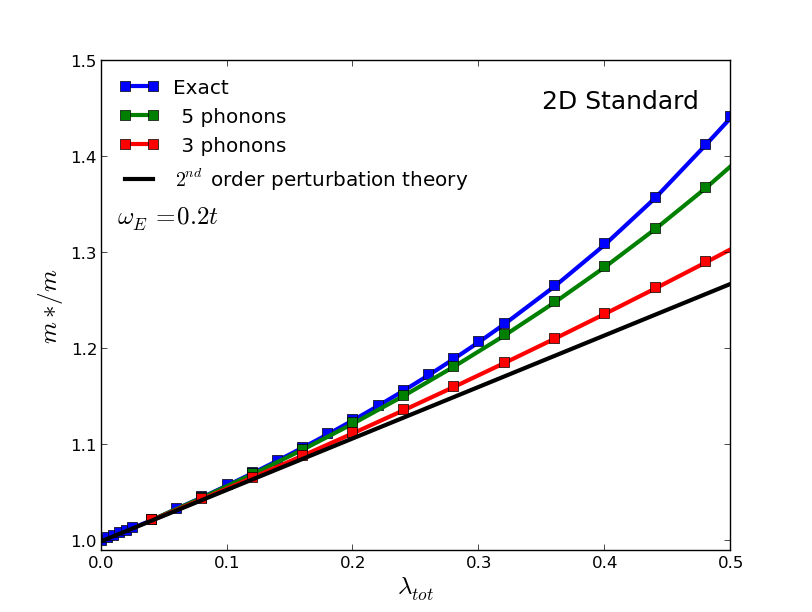}
\caption{ $m^\ast/m$ decreases as the accuracy of the calculation decreases. The curves with 5 and 3 phonons include 
states with phonons far from the electron, but never more than 5 or 3 phonons total respectively. 2nd order perturbation
theory includes excited states with at most a single phonon. $\omega_E = 0.2t$}
\label{fig:meff_N_phonon_2D_ordinary}
\end{center}
\end{figure}

\begin{figure}
\begin{center}
\includegraphics[height=3.5in,width=3.5in]{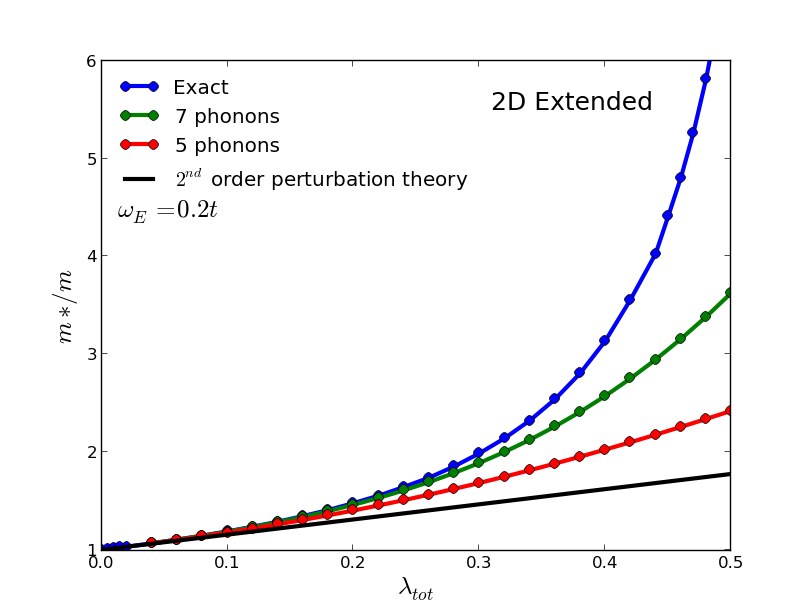}
\caption{For the 2D extended Holstein model, $m^\ast/m$ again decreases as the accuracy of the calculation decreases similar to the standard model. Here the
effect is even more pronounced: many phonons are required even for moderate coupling strengths. Again, 2nd order perturbation
theory includes at most a single phonon. $\omega_E = 0.2t$}
\label{fig:meff_N_phonon_2D}
\end{center}
\end{figure}
\end{subfigures}

\begin{figure}
\begin{center}
\includegraphics[height=3.7in,width=3.7in]{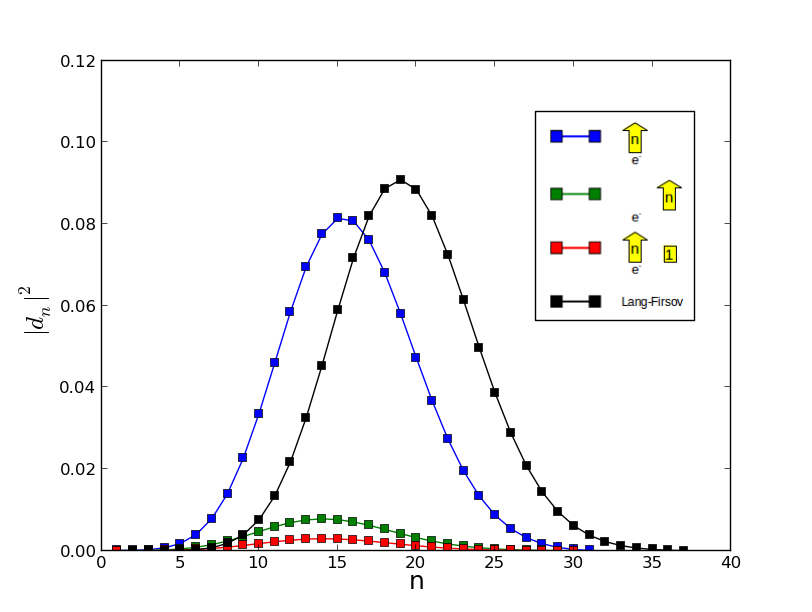}
\caption{ Probability ($|d_n|^2$) of various basis states from the Lang-Firsov approximation and 
the standard model in 2D,  with $\lambda = 0.62$ and $\omega_E = 0.2t$. The legend shows the distribution of 
phonons around the electrons for each `coherent state', while the Lang-Firsov wavefunction only has phonons on the same site as 
the electron. Lines are a guide to the eye only.}
\label{poisson}
\end{center}
\end{figure}

A second point we wish to make concerning the standard model is that, contrary to what the (fairly accurate) results for
the ground state energy might imply, the strong coupling solution is not simply the single Lang-Firsov coherent state:
\begin{equation}
|\psi\rangle = e^{-g^2/2}{1 \over \sqrt{N}} \sum_{\ell} e^{ikR_\ell} e^{g\hat{a}^\dagger_\ell}\hat{c}^\dagger_\ell |0\rangle.
\label{coherent}
\end{equation}
\noindent We have done an exact calculation and compare the resulting exact wavefunction to the Lang-Firsov coherent state
in Fig. \ref{poisson}. The exact solution has a large coherent state component which has a peak somewhat shifted from
the Lang-Firsov coherent state. The really important difference is the contribution of states with phonons that are 
not on the same site as the electron. While these states do not have a large probability in the overall wavefunction, they are important for calculating an accurate effective mass. These states increase the size of the phonon cloud
so that neighboring sites are ``more prepared'' to receive the electron and thus lower the overall effective mass. 
Using only the Lang-Firsov coherent state gives an effective mass that is too high by orders of magnitude. The same 
principle reduces the effective mass in the extended model; a nearest neighbour coupling produces phonons on the 
neighbouring sites, preparing them to receive the electron and lowering the effective mass.  

As illustrated in Fig. \ref{poisson} the crucial addition of extended phonons can be grouped with great precision
into multiple coherent states, and coherent states modified with a few other phonons.  So, in general, the wave
function could be expanded into multiple coherent components:
\begin{eqnarray}
|\psi\rangle &\approx& b_0 e^{-g_0^2/2}{1 \over \sqrt{N}} \sum_{\ell} e^{ikR_\ell} e^{g_0\hat{a}^\dagger_\ell}\hat{c}^\dagger_\ell |0\rangle
\nonumber \\
&+& b_1 e^{-g_1^2/2}{1 \over \sqrt{4N}} \sum_{{\ell},\delta} e^{ikR_\ell} e^{g_1\hat{a}^\dagger_{\ell + \delta}}\hat{c}^\dagger_\ell |0\rangle
\nonumber \\
&+& b_2 e^{-g_2^2/2}{1 \over \sqrt{4N}} \sum_{\ell, \delta} e^{ikR_\ell} e^{g_2\hat{a}^\dagger_\ell}\hat{a}^\dagger_{\ell + \delta}
\hat{c}^\dagger_\ell |0\rangle + ...,
\label{many_coherent}
\end{eqnarray}
\noindent where $\delta$ designates neighbouring sites in all directions. Based on how orderly Fig.~\ref{poisson}
looks one could imagine using a variational approach with these coherent states as well.
For this paper, however, we kept with the simple Bloch states which though far more numerous are 
easier to handle as they are guaranteed to be orthogonal:
\begin{equation}
|\psi \rangle  = \sum_n d_n \left( \frac{1}{\sqrt{N}} \sum_{\ell} e^{ikR_\ell} |\phi_{n \ell} \rangle \right),
\end{equation}
\noindent where the $|\phi_{n \ell} \rangle$ are orthonormal product states consisting of and an electron at site $\ell$ and phonons at sites $ \ell + \delta$. For example, for the Lang-Firsov state given by Eq. (\ref{coherent}), $d_n = e^{-g^2/2}g^n/\sqrt{n!}$.

The main message of this is that while the standard Holstein model may be the simplest polaron model, it's solution
is still a fairly complicated many body wavefunction, even in the weak and strong coupling limits. 
Without accurate characterization of this 
wavefunction it may still be possible to calculate some expectation values, like the ground state energy, accurately, 
but others, such as the effective mass, need a more precise wavefunction. When 
measuring expectation values with any numerical method, it is important to converge these 
values on their own as they may need a much larger basis space than, for example, the ground state energy, to be
accurately described.

\begin{figure}
\begin{center}
\includegraphics[height=3.5in,width=3.5in]{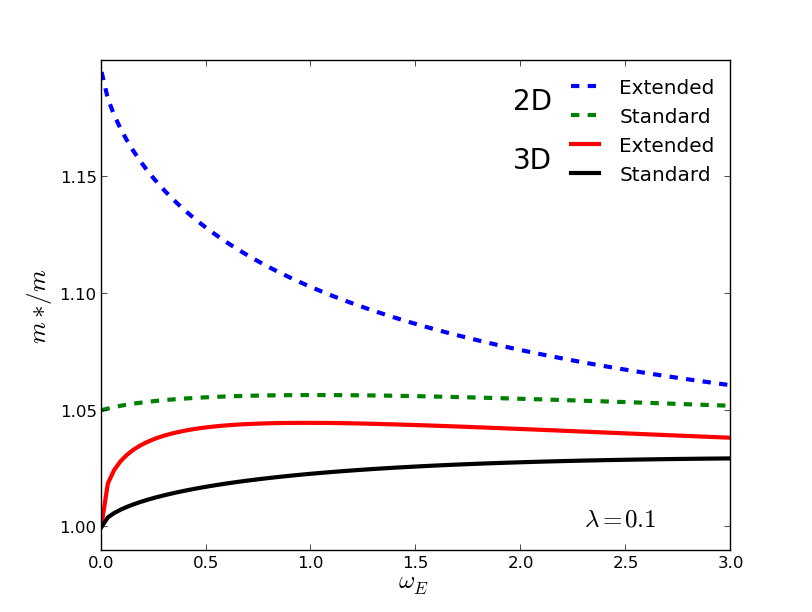}
\caption{  $m^\ast/m$ from 2nd order perturbation theory vs $\omega_E$. For definiteness, the value 
of $\lambda_{\rm tot}$ is $0.1$.
In the large $\omega_E$ limit all the curves approach $1.0$ but the Extended model always has a larger $m^\ast/m$ than the Standard 
model in both 2D and 3D. }
\label{fig:MeffVsOmega_23D}
\end{center}
\end{figure}

The first qualitative difference found between the extended and on-site Holstein models
is at weak coupling. In the standard model, we find a very slight increase of the effective mass with increasing phonon frequency
(see Fig.~\ref{fig:MeffVsOmega_23D}), while in the extended model, we find the opposite; the effective mass decreases
as a function of increasing frequency. Note that the analytical result for the effective mass in the adiabatic limit
is obtained from perturbation theory, and not from the semi-classical adiabatic calculations.\cite{kabanov93} The
adiabatic calculation reveals in weak coupling a regime in which there are {\it no} ion deformations, possibly indicating an
effective mass equal to the non-interacting electron mass. This is incorrect, and perturbation theory calculations correctly
yield an effective mass ratio equal to ($1 + \lambda/2$) for the standard Holstein model in two dimensions. 
For the extended model used here, one can show that the perturbative effective mass is given by
\begin{equation}
m^\ast/m = 1 + {\lambda_{\rm tot} \over 2} {[1 + 4f(1)]^2 \over 1 + 4 f^2(1)}.
\label{pert_adiab_extended}
\end{equation}
\noindent This agrees with the limiting value as $\omega_E \rightarrow 0$, obtained through numerical integration 
in  Fig.~\ref{fig:MeffVsOmega_23D}.

Since the exact results and perturbation theory agree for the effective mass in very weak coupling
the latter calculations can be trusted in this regime. Therefore we plot in Fig.~\ref{fig:MeffVsOmega_23D} 
only the perturbation theory results for the effective mass vs. $\omega_E/t$ to 
highlight the differences between the extended and standard models (for definiteness, we use $\lambda_{\rm tot} = 0.1$).
At low values of $\omega_E/t$ the extended model's effective mass decreases 
monotonically with increasing $\omega_E/t$ while the on-site
model has a peak near $\omega_E/t = 1$ . Both models have an effective mass ratio of unity ($m^\ast/m = 1$) 
in the anti-adiabatic limit, $\omega_E \rightarrow \infty$

%
%
%

%

Quantitatively, the extended Holstein model has a larger effective mass at weak coupling, and a smaller
effective mass at strong coupling compared to the standard Holstein model (for the same phonon frequency). 
Previous workers\cite{marsiglio93,alexandrov99,bonca99} have examined both the ordinary and/or extended models
in the past, but due to computational considerations did not examine the model in the physically 
most important parameter regime. In many real materials the electronic bandwidth is in the eV range while
the phonon modes tend to be in the meV range, so the physical adiabatic ratio regime is 
$\omega_E/t {{\atop <}\atop {\approx \atop} } 0.2$.

\begin{subfigures}
\begin{figure}
\begin{center}
\includegraphics[height=3.5in,width=3.5in]{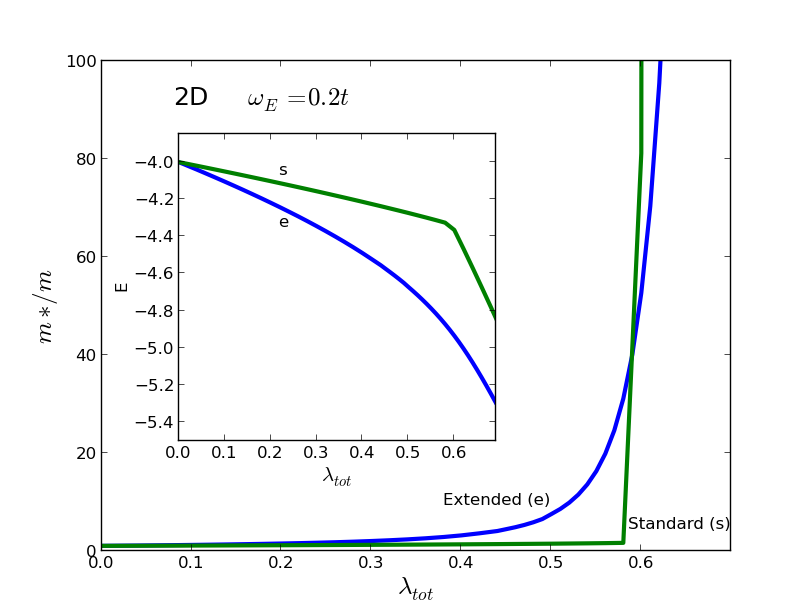}
\caption{ For the two dimensional model, $m^\ast/m$ as a function of $\lambda_{\rm tot}$ up to strong coupling with $\omega_E =0.2t$. Note that beyond 
$\lambda_{rm tot} = 0.6$ the effective mass of both models is practically infinite, but at intermediate coupling
strengths the extended model shows a significantly higher effective mass. {\it Inset:} Ground state energy vs. $\lambda_{\rm tot}$.}
\label{fig:meff_strongCouplingLowW_2D}
\end{center}
\end{figure}
\begin{figure}
\begin{center}
\includegraphics[height=3.5in,width=3.5in]{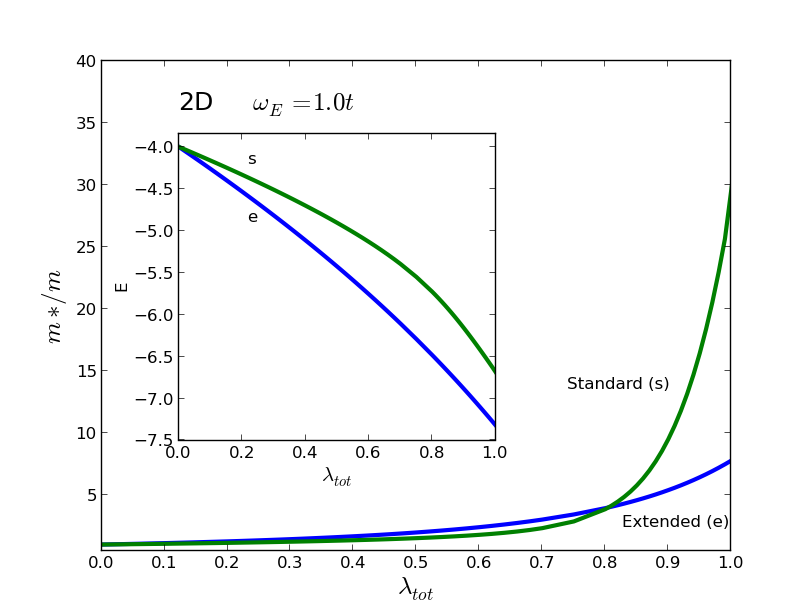}
\caption{ For the two dimensional model, $m^\ast/m$ as a function of $\lambda_{\rm tot}$ up to strong coupling 
with $\omega_E =t $. Note that the 
effective masses are all much smaller and at intermediate coupling
strengths the extended model shows an effective mass only slightly higher than the standard model. The coupling
strength is also much larger before the effective masses rise to unphysical values. {\it Inset:} Ground state energy 
vs. $\lambda_{\rm tot}$.}
\label{fig:meff_strongCouplingHighW_2D}
\end{center}
\end{figure}
\end{subfigures}

For reasons stated earlier, calculations with small phonon frequencies are difficult to converge, so we
have utilized a number of small frequencies. As a compromise, we will present most results for $\omega_E/t = 0.2$, as these
are well converged, and also representative of the low frequency regime. In Figs. \ref{fig:meff_strongCouplingLowW_2D}
and \ref{fig:meff_strongCouplingHighW_2D} we show the effective mass ratio vs coupling strength for $\omega_E/t = 0.2$
and $\omega_E/t = 1.0$, respectively, for the standard vs extended Holstein models. Several points should be made. 
First, the crossover from the so-called weak to the so-called strong coupling regime for both standard and extended
models is not so clear when $\omega_E/t \approx 1$, especially for the extended model. In contrast, the crossover is
much better delineated for low phonon frequencies, as is clear in Fig. \ref{fig:meff_strongCouplingLowW_2D}. The point made
in Ref. [\onlinecite{alexandrov99}], that the extended model results in a lower effective mass is also clear, for $\lambda$ values
beyond some intermediate coupling strength. At high phonon frequency (Fig. \ref{fig:meff_strongCouplingHighW_2D}) 
this means a reduction from $m^\ast/m \approx 30$ to $m^\ast/m \approx 7$, for example. By any measure this reduction is
significant, but somewhat irrelevant, since the effective masses involved, even after reduction, 
are too high to describe normal state properties. However, for more realistic (lower) phonon frequencies (Fig. \ref{fig:meff_strongCouplingLowW_2D}), both
crossovers are sharpened as a function of coupling strength, although less so for the extended model, with the net result that
a regime of effective mass reduction remains, and the reduction is enormous, but now the mass is `lowered' to values of
$40$ or higher. In fact, the clear effect of extended interactions is to {\it raise} the effective mass for most of the parameter
regime that is physically relevant.\cite{remark3} 

To summarize this subsection, the extended Holstein model is more realistic than the standard model insofar
as it includes longer range interactions. This does give rise to a coupling regime where the effective mass is lowered,
compared to the standard model, but we argue that lowering the effective mass ratio from $100$ to $40$ is not so relevant. 
Instead, the clear result of increasing the range of interaction is to {\it enhance}
the effective mass so that in the so-called weak coupling regime the effective mass is increased due to the longer range
interactions. Extended range interactions is therefore {\it not} seen as a means to lower the electron effective mass
to reasonable levels in the 2D polaron problem. However, the effect of long range interactions can be different in three
dimensions, and we turn to that question next.
 
\section{Results in 3D} 

\begin{figure}
\begin{center}
\includegraphics[height=3.7in,width=3.7in]{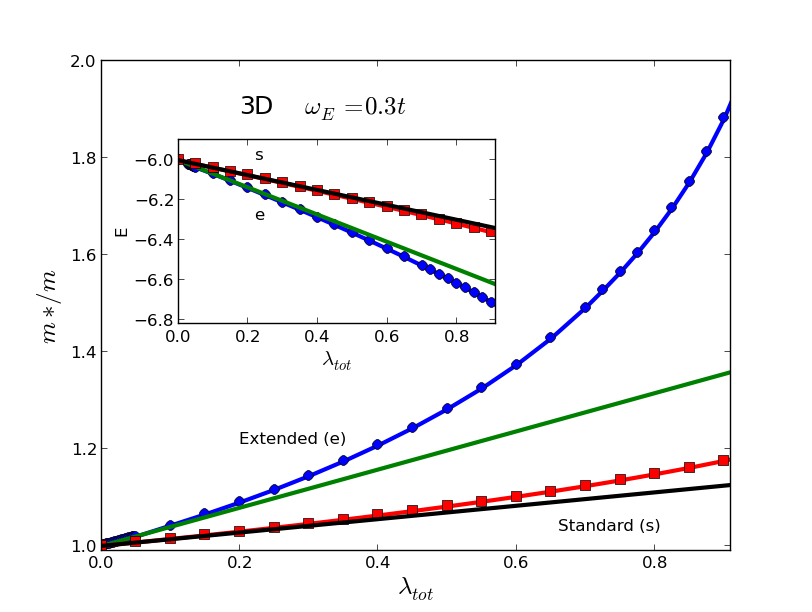}
\caption{For the three dimensional model, $m^\ast/m$ as a function of $\lambda_{\rm tot}$, with $\omega_E =0.3t$,
to compare exact and perturbative results. 
Similar to the 2D results the region of validity for perturbation theory is much smaller in the extended 
model though it is not particularly large for the standard model either. {\it Inset:} Ground state energy vs. $\lambda_{\rm tot}$. 
Again, perturbation theory works better for the ground state energy than for the effective mass. }
\label{fig:meff_perturbation_3D}
\end{center}
\end{figure}

We applied the same technique to an extended Holstein model in three dimensions. We 
limited ourselves again to nearest neighbour and on site electron phonon coupling, simply 
extending the two-dimensional Hamiltonian to three dimensions. We then
then computed the ground state energy and effective mass with perturbation theory
and with the numerical method outlined above.

\begin{figure}
\begin{center}
\includegraphics[height=3.7in,width=3.7in]{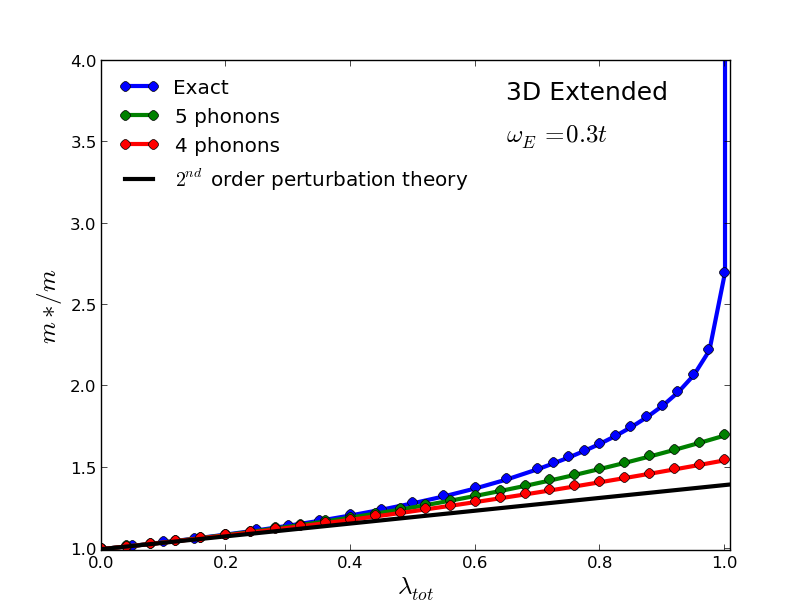}
\caption{For the three dimensional case, $m^\ast/m$ vs. $\lambda_{tot}$, with $\omega_E = 0.3t$. 
The effective mass increases as more and more states with more than one phonon excitation are included.
The curves with 4 and 5 phonons include 
states with phonons far from the electron, but there are never more than a total of 4 or 5 phonons, respectively. The
calculation with 2nd order perturbation theory includes basis states with at most a single phonon.}
\label{fig:meff_perturbation_3D_converge}
\end{center}
\end{figure}

The perturbation theory was done with straightforward Rayleigh-Schr\"{o}dinger perturbation theory
with the integrals performed numerically as in the 2D case. We used $\omega_E/t = 0.3$, since 
the electronic bandwidth with nearest neighbour hopping only is $12t$ in 3D (as opposed to $8t$ in 2D). 
As in the 2D case there is good agreement between the exact results and perturbation theory 
only for a small region of weak coupling, $\lambda_{\rm tot} {{\atop <}\atop {\sim \atop} } 0.4$, for the standard model,
as seen in Fig.~\ref{fig:meff_perturbation_3D}. In the extended model, the good agreement is only achieved for
an even more restricted range of $\lambda_{\rm tot} {{\atop <}\atop {\sim \atop} } 0.2$, as seen in the same figure.

\begin{figure}
\begin{center}
\includegraphics[height=3.7in,width=3.7in]{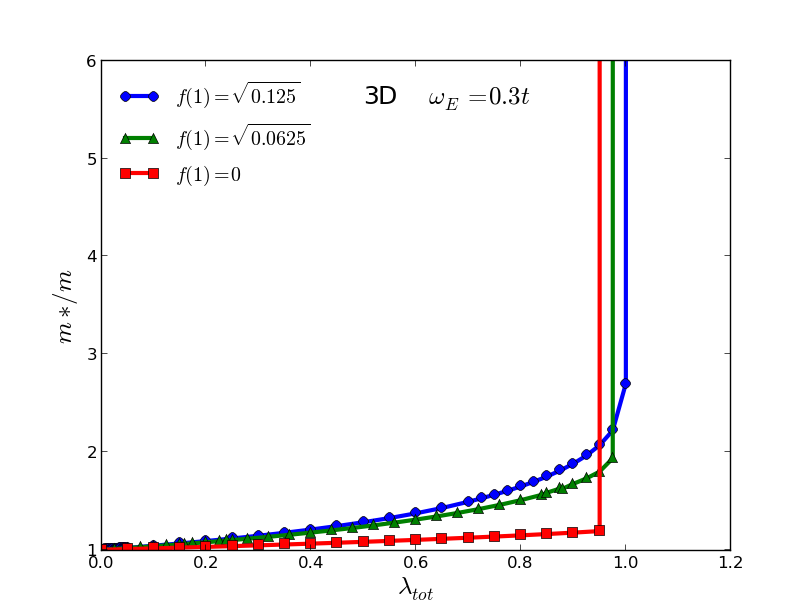}
\caption{For the three dimensional case, $m^\ast/m$ vs. $\lambda_{tot}$ with various values of $f(1)$. These
results are from numerically exact calculations. 
The extended model can shift the onset of unphysically high effective masses to stronger coupling, though the extra
range of tenable coupling strengths is rather small.}
\label{fig:meff_exact_3D}
\end{center}
\end{figure}

These results are plotted for a `low' phonon frequency, $\omega_E/t \approx 0.3$; however, in 3D there  
is no ambiguity about the adiabatic limit. Perturbation theory, semi-classical adiabatic calculations\cite{kabanov93}
and limiting trends from exact diagonalization all agree that for small $\lambda$, the effective mass ratio
approaches unity, i.e. $m^\ast/m \rightarrow 1$  as $\omega_E \rightarrow 0$, and there is no polaron formation in
this limit. In Fig.~\ref{fig:MeffVsOmega_23D} it is clear from the 3D results that the behaviour of the effective mass ratio
for the extended model is quite similar to that for the standard Holstein model. Quantitatively, the effective mass is somewhat
larger for the extended model, but not enormously so. The results in Fig.~\ref{fig:meff_perturbation_3D}
are both representative of the adiabatic limit, which has no polaron formation.


In Fig.~\ref{fig:meff_perturbation_3D_converge} we show the so-called perturbative regime to illustrate that
also in three dimensions this regime is confined to very small values of $\lambda$ only. The crossover region,
shown for the effective mass in  Fig.~\ref{fig:meff_exact_3D},
becomes very sharp as the phonon frequency is decreased,\cite{ku02,li12} to coincide with the point
in semiclassical calculations where the ground state abruptly becomes polaronic, after being
free-electron-like up to that point. The impact of longer range interactions is clear from the figure; the
crossover region is definitely moved to higher coupling strengths as the range and strength of the nearest
neighbour interaction increases. An obvious limiting case is where the interaction becomes infinitely long
ranged, with the same strength independent of distance from the electron. In this case the electron will remain
free-electron-like for all coupling strengths. Note that for realistic nearest neighbour interactions there is now a 
small range of coupling strengths where the effective mass is indeed reduced to realistic values through longer-range 
interactions, consistent with the original conclusions of Ref. [\onlinecite{alexandrov99}]. That is, in contrast
to the two-dimensional case, extended range interactions seem to shift the regime of coupling strengths wherein
the effective mass is low, {\it without at the same time increasing significantly the effective mass in this regime.}

\section{Summary} 

We have presented exact and perturbative results for the extended Holstein model. This model was conceived\cite{alexandrov99}
in an effort to realize the Fr\"ohlich model on a lattice. Our primary purpose was to re-assess the conclusions of
Ref. [\onlinecite{alexandrov99}] when a smaller and more realistic adiabatic ratio, $\omega_E/t$, is used. We found that, in two dimensions,
the effective mass in the so-called weak coupling regime is enhanced by longer range interactions, while the effective mass
in strong coupling is suppressed. This is in agreement with the results of  Ref. [\onlinecite{alexandrov99}], but we nonetheless
find this assessment misleading. In particular, in strong coupling, being able to achieve an effective mass reduction from $100$
to $40$ is wonderful, but does not serve to reconcile the qualitative results of the Migdal description with the single polaron 
results. It still remains that, over most of the range of coupling strengths, the effective mass is {\it increased} by longer range
interactions. Moreover, it is clear that a perturbative description, which, for a single electron problem actually coincides
(in a technical sense) with the Migdal approximation, is woefully inaccurate when it comes to describing the details (wave function,
electron effective mass, etc.) of the solution, even for much weaker coupling strengths. 
In three dimensions this problem is slightly ameliorated, in that, at least for a very
limited range of coupling strengths, the effective mass can be vastly reduced by many orders of magnitude by longer range
interactions (see Fig.~\ref{fig:meff_exact_3D}, just to the right of the standard Holstein results, i.e. the left-most almost vertical line).
Even in three dimensions, however, the perturbative weak coupling regime seems to require considerably more phonon
excitations then second order perturbation theory would suggest.

Further attempts to reconcile Migdal-based approximations versus exact single electron calculations can proceed along
a number of paths, several of which are currently under investigation. For example, one can attempt to develop controlled
approximations for more than one electron (the bipolaron, for example, has been already
investigated).\cite{bonca01,davenport12}
One would expect phonon and coupling strength renormalization to occur as the increasing number of electrons will
have a more significant impact on the phonons. Another direction involves more sophisticated electron phonon couplings, and
the possible importance of acoustic modes.

\begin{acknowledgments}
This work was supported in part by the Natural Sciences and Engineering Research Council of Canada (NSERC) and by
the Alberta Innovates Techology Futures (AITF) program.

\end{acknowledgments}


\begin{thebibliography}{99}
\bibitem{remark3} This fact was also pointed out in Ref. [\onlinecite{alexandrov99}] --- see their Fig. 4, for example,
but was de-emphasized in their work in favour of the effective mass decrease at higher coupling strengths.
\end{thebibliography}
\end{document}